\documentclass[%
 reprint,
 amsmath,amssymb,
 aps,
]{revtex4-1}

\usepackage[pdftex]{graphicx}

\newcommand{\eq}[1]{(\ref{#1})}
\newcommand{\la}[1]{\label{#1}}

\newcommand{\ckappaa}[2]{\Big(\!\!\!\Big({\color{MidnightBlue}#2}\Big)\!\!\!\Big)_{\!#1}}

\def\[{\left[}
\def\]{\right]}
\def\({\left(}
\def\){\right)}
\def\[{\left[}
\def\]{\right]}

\def\<{\langle}
\def\>{\rangle}

\usepackage[usenames,dvipsnames]{xcolor}
\usepackage{color}
\usepackage{graphicx}
\usepackage{dcolumn}
\usepackage{bm}

\newcommand{\beq}{\begin{equation}}
\newcommand{\eeq}{\end{equation}}
\newcommand{\beqq}{\begin{equation*}}
\newcommand{\eeqq}{\end{equation*}}
\newcommand\beqa{\begin{eqnarray}}
\newcommand\eeqa{\end{eqnarray}}
\newcommand\beqaa{\begin{eqnarray*}}
\newcommand\eeqaa{\end{eqnarray*}}
\newcommand\bea{\begin{array}}
\newcommand\eea{\end{array}}

\begin{document}

\preprint{APS/123-QED}

\title{Quantum Integrability for Three-Point Functions}

\author{Nikolay Gromov}
\affiliation{%
King's College London, Department of Mathematics,
London WC2R 2LS, UK \\
%
\& PNPI,
Gatchina, 188 300, St.Petersburg, Russia}
\author{Pedro Vieira}%
\affiliation{%
Perimeter Institute for Theoretical Physics,
Waterloo, Ontario N2L 2Y5, Canada
}%


\begin{abstract}
Quantum corrections to three-point functions of scalar single trace operators in planar ${\cal N}=4$ Super-Yang-Mills theory are studied using integrability. At one loop, we find  new algebraic
 structures that not only govern all two loop corrections to the mixing of the operators but also automatically incorporate all one loop diagrams correcting the tree level Wick contractions.
Speculations about possible extensions of our construction to all loop orders are given. We also match our results with the strong coupling predictions in the classical (Frolov-Tseytlin) limit.

\end{abstract}

\pacs{Valid PACS appear here}
\maketitle


\section{\label{sec:intro}Introduction}

In this paper we consider three-point correlation functions (3pt CF) of single trace gauge invariant operators of planar $\mathcal{N}=4$ supersymmetric Yang-Mills (SYM) theory.
We consider mostly the first quantum correction (one loop) to the leading result (tree level) of \cite{paper1} and speculate about some all loop features at the very end. The motivation for this study is twofold.
On the one hand, the knowledge of the spectrum \cite{review} together with the 3pt CF will suffice to determine any correlation function in this $3+1$ dimensional quantum field theory in a completely non-perturbative fashion.
Such a highly ambitious goal is believed to be attainable due to the \textit{Integrability}, or exact solvability, of planar $\mathcal{N}=4$ SYM \cite{review}.
 Another motivation
is to better understand Holography and the emergence of a dual string description of a quantum gauge theory.
How do smooth string worldsheets come about? Do they have a natural Integrable description in $\mathcal{N}=4$ SYM? 3pt CF might be a great playground for addressing some of these questions. In particular, as we will reinforce in this letter, the answer to the last question seems to be \textit{yes}; three-point functions can be studied most efficiently using integrablity.
\begin{figure}[t]
\includegraphics[scale=.38]{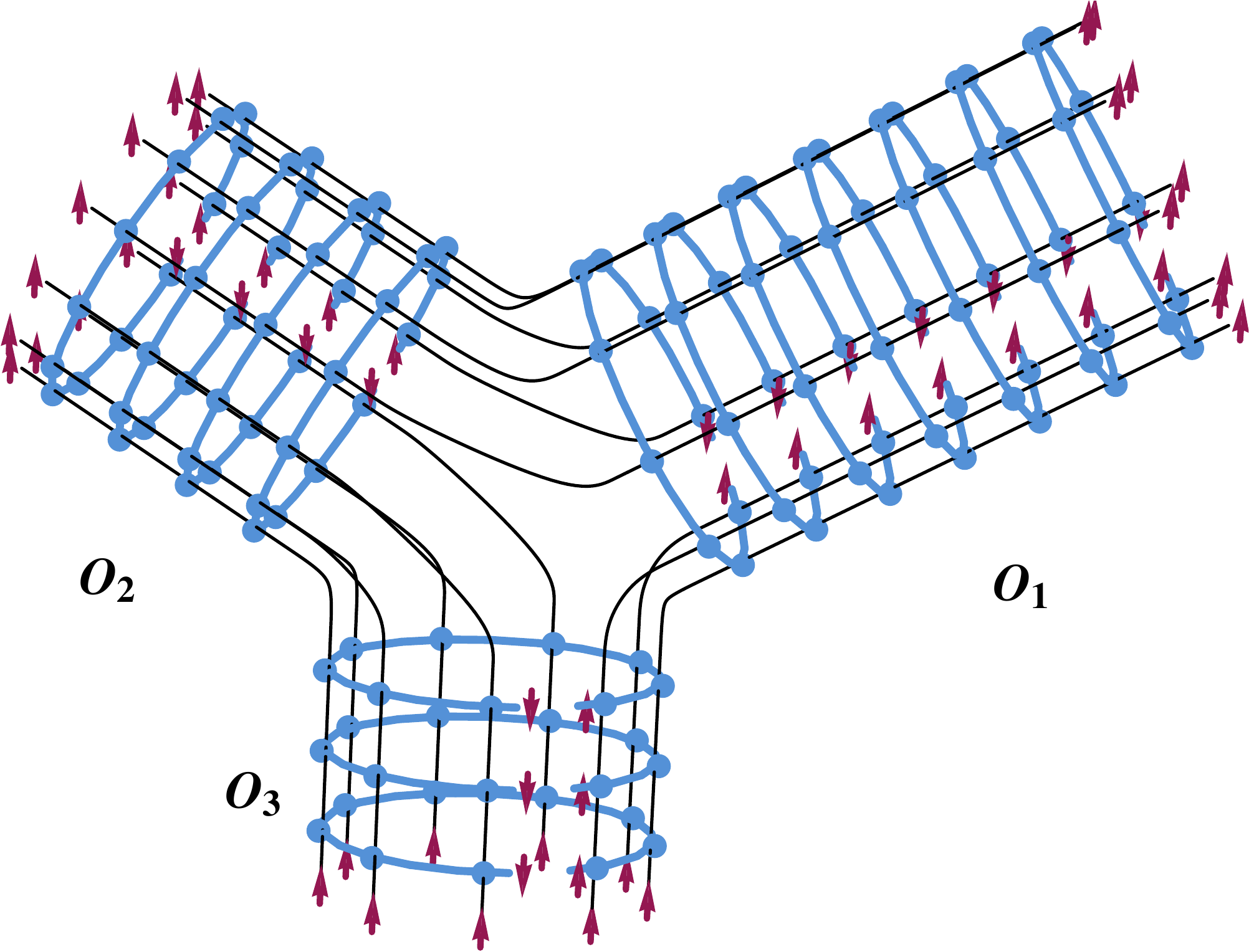}
\caption{\label{fig:epsart} Tree level correlation function of three single trace operators. Each operator $\mathcal{O}_i$ is obtained by acting on a  vacuum with a set of $N_i$ creation operators (blue thick lines). This generates a state with $L_i$  spins (thin black lines), $N_i$ of which are flipped. These states are then glued together. We end up with a vertex model partition function with the topology of a thrice punctured sphere; it strongly resembles a discrete string path integral.
We have $N_1=N_2+N_3$ so all spins $\downarrow$ from $\mathcal{O}_2$ and $\mathcal{O}_3$ are contracted with $\mathcal{O}_1$. Since there are $N_3$ thin lines connecting $\mathcal{O}_1$ and $\mathcal{O}_3$ all those lines are $\downarrow$ spins; see \cite{paper1} for details.
 } \label{sphere}
\end{figure}
\section{Two Loop eigenstates}
To compute the correlation functions at one loop we need to solve the two loop mixing problem. This is the subject of the current section. As in \cite{paper1}, we consider operators made out of two complex scalars (which are identified with states with $\uparrow$ and $\downarrow$ spins) that diagonalize the dilatation operator \cite{BKS}
\begin{equation}
\hat H =  (2g^2-8g^4) \sum_{i=1}^L \mathbb{H}_{i,i+1} + 2 g^4 \sum_{i=1}^L \mathbb{H}_{i,i+2} +\mathcal{O}(g^6)\,. \label{Heq}
\end{equation}
Here $\mathbb{H}_{a,b} \equiv \mathbb{I}-\mathbb{P}_{a,b}$ with $\mathbb{P}$ being the permutation operator and sites $L+1$ and $1$ are identified.  The fundamental excitations are magnons (spins $\downarrow$) moving in a ferromagnetic vacuum (where all spins are $\uparrow$). Their energy and momentum are parametrized as $ E(u)= 2i g^2\left({1}/{x^+}-{1}/{x^-}\right)$ and $p(u)=i \log({x^-}/{x^+})$
where the Zhukowsky variables $x^{\pm}=(u\pm i/2) - g^2/(u\pm i/2)+\mathcal{O}(g^4)$.
The simplest state diagonalizing \eq{Heq} is the single magnon
\begin{equation}
\sum_{n=1}^L \left(\frac{x^+}{x^-} \right)^{n}  |\underbrace{\uparrow\dots \uparrow }_{n-1} \downarrow\uparrow \dots \uparrow \rangle  \label{coordinate} \,.
\end{equation}
At leading order in perturbation theory, there is an equivalent description of the states using the algebraic Bethe ansatz formalism (see \cite{paper1} for a review). E.g., the single magnon state (\ref{coordinate}) simplifies to  
\begin{equation}
\sum_n \left(\frac{u+i/2}{u-i/2} \right)^{n}  \sigma_n^- \left|\uparrow\dots \uparrow\right\rangle  \propto \hat B(u) \left|\uparrow\dots \uparrow\right\rangle  \label{algebraic}
\end{equation}
where the creation operators are given by
\begin{equation}
{\includegraphics[scale=0.45]{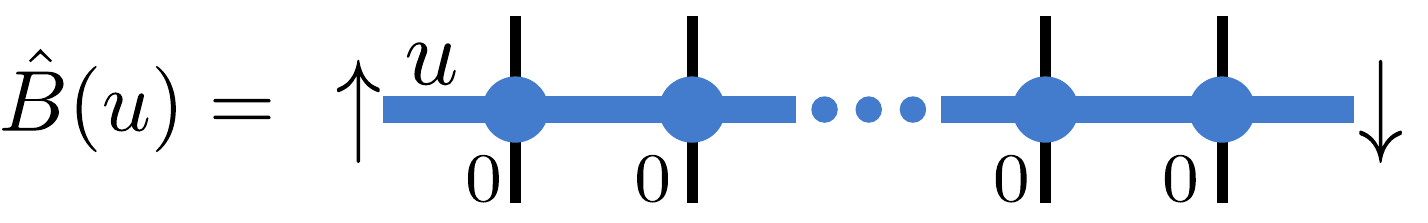} }
\label{Beq} \end{equation}
with the R-matrix given by
\begin{equation}
{\includegraphics[scale=0.43]{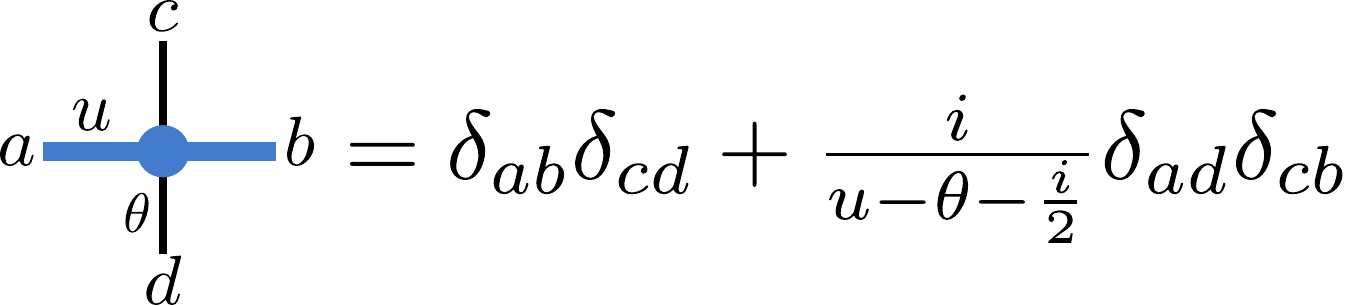} } \nonumber
\end{equation}
The algebraic treatment reveals its elegance when we consider states with $N$ \textit{interacting} magnons. This multi-particle state is simply given by
\begin{equation}
\hat B(u_1)\dots \hat B(u_N) \left|\uparrow\dots \uparrow\right\rangle \,.   \label{Bleading}
\end{equation}
Each of the legs in figure \ref{sphere} corresponds to one such state.
The energy of these states is given by $\sum E(u_i)=2g^2 \Gamma_{\bf u}$,
\begin{equation}
\Gamma_{\bf u}=\sum_{i=1}^N \frac{1}{u_i^2+\frac{1}{4}} +\mathcal{O}(g^2)\,. \nonumber
\end{equation}
At tree level we should contract the states as in fig. \ref{sphere} \cite{paper1}.

At the next loop order we need to improve (\ref{Bleading}) to obtain the two loop spin chain eigenstates. There are no explicit expressions for these eigenstates in the literature. We will now describe how to
construct them using a modification of the algebraic Bethe ansatz. From (\ref{algebraic}) we see that we want to modify the propagation of the magnon along the chain to get the correct dispersion relation. The simplest way to achieve this preserving integrability is to introduce impurities $\theta_j$ at each site converting (\ref{Beq}) into
\begin{equation}
\includegraphics[scale=.50]{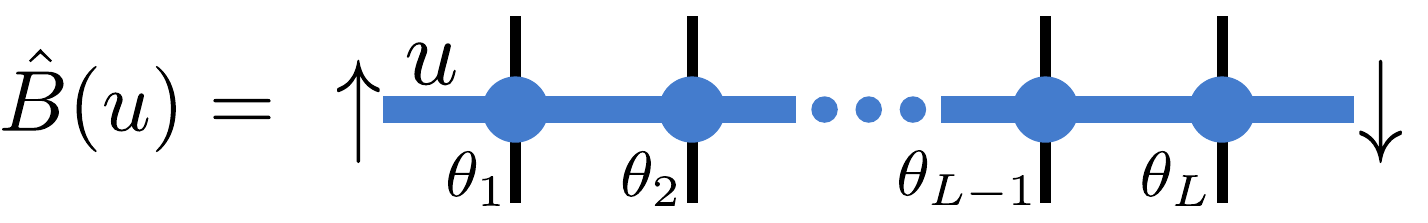} \label{Btheta}
\end{equation}
With these modified creation operators, the single magnon state $ \hat B(u) \left|\uparrow\dots \uparrow\right\rangle  $ takes the form
\begin{equation}
\sum_{n=1}^L \left[\prod_{k=1}^{n-1} \frac{u-\theta_k+\frac{i}{2}}{u-\theta_k-\frac{i}{2}}\right] \frac{i}{u-\theta_n-\frac{i}{2}}  |\underbrace{\uparrow\dots \uparrow }_{n-1} \downarrow\uparrow \dots \uparrow \rangle
\;.
 \label{coordinateImp}
\end{equation}
The idea is to use the impurities $\theta_k$ to realize the required correction to the dispersion which arises at two loops. To achieve this we introduce the differential operator
\begin{equation}
\ckappaa{\theta }{f} \equiv \left. f+\frac{g^2}{2} \sum_{i=1}^L \(\partial_{\theta_{i}}-\partial_{\theta_{i+1}}\)^2 f\right|_{\theta_j \to 0} \!\!\!\! +\mathcal{O}(g^4) \label{kappa}
\end{equation}
which we call the  \textit{$\Theta$-derivative}. Here $\partial_{L+1}$ is identified with $\partial_1$.
It is easy to verify that applying the $\Theta$-derivative to (\ref{coordinateImp}) we reproduce the good state (\ref{coordinate}) modulo a simple mismatch at the boundaries for the $n=1,L$ terms in  (\ref{coordinate}).
What is way more remarkable is that, not only that mismatch can be fixed, but in fact,
\begin{equation}
\(1-g^2\,\Gamma_{\bf u}\, \mathbb{H}_{L,1} \) \ckappaa{\theta}{\hat B(u_1)\dots \hat B(u_N)  \left|\uparrow\dots \uparrow\right\rangle } \label{promotion}
\end{equation}
yields perfect \textit{N-magnon} eigenstates of the two loop $\mathcal{N}=4$ SYM dilatation operator \cite{paper4}!

\section{3pt Functions with impurities}
The contractions between operators $\mathcal{O}_3$ and the other two operators are trivial, see caption of figure \ref{sphere}. The ones between $\mathcal{O}_3$ and $\mathcal{O}_2$ are simply contractions of $L_3-N_3$ $\uparrow$ spins while the contractions between $\mathcal{O}_3$ and $\mathcal{O}_1$ involve  $N_3$ $\downarrow$ spins. That is, the effect of the operator $\mathcal{O}_3$ is to remove a piece of ferromagnetic vacuum of length $L_3-N_3$ from $\mathcal{O}_2$ and replace it with a sequence of magnons of length $N_3$. In formulas, $\left| 2\right\>\equiv \hat B(v_1)\dots \hat B(v_{N_2}) \left|\uparrow\right\>^{\otimes L_2} \to \hat{\mathcal{O}}_3 |{2}\rangle$ where \cite{roiban}
\begin{equation}
\hat{\mathcal{O}}_3 =\( \left|\downarrow\right\>^{\otimes N_3}\)\Big( {}^{\otimes L_3-N_3}\left\< \uparrow\right| \Big)\,.\end{equation}
The operator $\hat{\mathcal{O}}_3 |{2}\rangle$, of length $L_1$, should be contracted with $\mathcal{O}_1$ given by $\left| 1\right\>\equiv\hat B(u_1)\dots \hat B(u_{N_1}) \left|\uparrow\right\>^{\otimes L_1}$. For simplicity, we will consider the case where the third operator $\mathcal{O}_3$ is a chiral primary. Then, the (absolute value of the properly normalized) tree level 3pt function with impurities is simply  \cite{paper1,Omar}
\begin{equation}
|{C}^{\rm tree\;with\;imp.}_{123}| = \frac{\sqrt{L_1L_2L_3}}{\sqrt{\binom{L_3}{N_3}}}  \, \frac{|\<1 | \hat{\mathcal{O}}_3| {2} \>|}{\sqrt{\left\<1 |1\right\>\left\<2 | 2 \right\>}} \label{main} \,.
\end{equation}
Let us specify which impurities we use in (\ref{Btheta}) when constructing $|1\>$ and $|2\>$.
Each thin line in figure \ref{sphere} has its own impurity. The impurities associated to the contractions between operator $\mathcal{O}_n$ and $\mathcal{O}_m$ are denoted by $\{\theta^{nm}_j\}$.
We define $\{\theta_j^{1} \}= \{\theta_j^{12} \} \cup  \{\theta_j^{13} \} $ etc.
Explicit expressions for the scalar products in (\ref{main}) are presented in the appendix. The tree level result
${C}^{\rm tree}_{123}$
in $\mathcal{N}=4$ SYM is given by (\ref{main}) if we send to zero all impurities. The impurities will be important when extending this expression to one loop.

\section{One loop 3pt Functions}
\begin{figure}[t]
\includegraphics[scale=1.2]{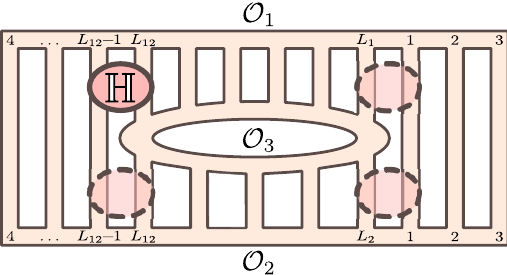}
\caption{
To take into account the loop diagrams correcting the Wick contractions
of the operators one must insert Hamiltonian densities at the junctions of the operators \cite{p1}.} \label{Hs}
\end{figure}

When computing 3pt CF at one loop, two effects need to be taken into account: (a) we need to correct the one loop operators into the two loop Bethe eigenstates and
(b) add insertions of Hamiltonians at the splitting points \cite{p1}. The first effect leads to (\ref{main}) where we replace the one loop
 states by the two loop eigenstates constructed via (\ref{promotion}) and indicated by boldface, 
\begin{equation}
|{C}^{\rm one\;loop\;(a)}_{123}| = \frac{\sqrt{L_1L_2L_3}}{\sqrt{\binom{L_3}{N_3}}}
\, \frac{|\<{\bf 1} |\hat{\mathcal{O}}_3 | {{\bf 2}} \>|}{\sqrt{\left\<{\bf 1} |{\bf 1}\right\>\left\<{\bf 2} | {\bf 2} \right\>}}  \la{ratio}\;.
\end{equation}
To compute this quantity we start with a tree level scalar product with impurities such as $\<1|1\>$. Then we act with the $\Theta$-derivative (\ref{kappa}) on it. When this differential operator acts on $|1\>$ we get $\bf |1\>$ up to a simple boundary term
 (\ref{promotion}). Same is true for $\<1|$. Then we also have the crossed terms when one of the derivatives in (\ref{kappa}) acts on $|1\>$ and another one acts on $\<1|$. These can be dealt with using
\begin{equation}
\left.i \(\partial_{\theta_j}-\partial_{\theta_{j+1}}\)\! \hat B(u) \right|_{\theta \to 0}\!\!\!= \! \[P_{j,j+1} + \delta_{j,L} \sum_{i=1}^L\mathbb{H}_{i,i+1}, \hat B(u)\]  \nonumber
\end{equation}
At the end of the day, we find \cite{paper4}
\begin{equation}
{\bf \left\<1 |1\right\>}= \[1-g^2 \(\Gamma^2_{\bf u}+2\Gamma_{\bf u}\)  \] \ckappaa{\theta^{1}}{\left\<1 |1\right\>}  \nonumber
\end{equation}
and an analogous expression for ${\bf \left\<2 |2\right\>}$.
Similarly, for the numerator, we find
\begin{eqnarray}
&&| { \<{\bf 1}  | \hat{\mathcal{O}}_3 | {\bf 2}\>} | \!=\!\Big|\[1-\!\tfrac{g^2}{2}\!\(\Gamma^2_{\bf u}+2\Gamma_{\bf u}+\Gamma^2_{\bf v}+2\Gamma_{\bf v}\!\)\]   \! \ckappaa{\theta^{1}}{\<1 | \hat{\mathcal{O}}_3| 2\> \!} \nonumber \\
&&\qquad+\,g^2 \,\<1|\mathbb{H}_{L_{12}-1,L_{12}} \hat{\mathcal{O}}_3|2\>+g^2\<1|\hat{\mathcal{O}}_3\mathbb{H}_{L_{12}-1,L_{12}} |2\> \nonumber\\
&&\qquad+\,g^2  \,\<1 |\mathbb{H}_{L_1,1} \hat{\mathcal{O}}_3\,\qquad|2\>+g^2\<1|\hat{\mathcal{O}}_3\mathbb{H}_{L_2,1} \qquad\,\, |  2\>
\Big |
\la{Slavnov}
\end{eqnarray}
where $L_{12}=L_1-N_3$. For the last two lines we should set the impurities to zero.
Two remarkable things happen when we put everything together. First, all the $\Gamma_{\bf u}$ and $\Gamma_{\bf v}$ cancel out when we construct the ratio (\ref{ratio}). Second, the last two lines in (\ref{Slavnov}) are nothing but Hamiltonian insertions at the splitting points
(see figure \ref{Hs}). They cancel \textit{precisely} with the Hamiltonian insertions which come from adding up all Feynman diagrams correcting the tree level Wick contractions \cite{p1}!
 As such,
when the dust settles, we end up with our main result
\beq
\left|{C_{123}^{\rm one\;loop}} \right| = \frac{\sqrt{L_1L_2L_3}}{\sqrt{\binom{L_3}{N_3}}}\, \frac{ \left|
\ckappaa{\theta^{1}}{\<1 | \hat{\mathcal{O}}_3|{2} \>}\right|}{\ckappaa{\theta^{1} }{\sqrt{\left\<1 |1\right\>}}\ckappaa{\theta^{2}}{\sqrt{\left\<2 | 2 \right\>}}} \label{eleven}
\eeq
for the structure constants up to one loop \cite{foot}. The striking simplicity of this result signals a deeper structure which the $\Theta$-derivative starts to unveil. The
derivatives in (\ref{eleven}) can
be explicitly computed with ease \cite{paper4}.

\section{Comparison with string theory}
The strong coupling regime of ${\cal N}=4$ SYM theory is described by classical strings.
Our results are, strictly speaking, valid at weak coupling. Yet, we shall demonstrate
that in a particular limit they coincide precisely with the string theory results.

The limit where one can in principle expect a match is the Frolov-Tseytlin limit \cite{FT}.  This is the limit of large operators $L_i\sim N_i\to\infty$ but with $g/L_i \ll 1$.
We will use the results of
 \cite{Costa} where $O_1\simeq O_2^\dagger$ correspond to two similar classical strings while $O_3$ is a small BPS string. The closest we can get to the Frolov-Tseytlin limit for all operators is then
\beq
1\ll N_3,L_3 \ll g\ll L_1,L_2,N_1,N_2 \,.
\eeq
This is the limit we consider.
As in \cite{paper2}, we will use the $SU(2)$ folded string solution
which is simple enough to work with and
has a rich structure at the same time. We also take $L_3=2N_3$ for the small operator $\mathcal{O}_3$. The result is then a function of  $3$ parameters only: $\alpha\equiv N_1/L_1$,  $L_1$ and $N_3$. The tree level weak coupling result  matches the leading order expansion in $g/L_1$ of the string theory result, denoted as
$C_{123}^{\rm tree}$ \cite{paper2} (see also \cite{strange}). For the next order, we find
%
%
\begin{equation}\la{strongpredic}
\frac{\!\!\!\!C_{123}^{\rm string}\!\!\!\!\!\!\!}{C_{123}^{\rm tree}}\!\simeq 1+ \frac{g^2 N_3}{L_1^2}\! \[ \frac{32 \alpha  (1-2 q) E^2(q)}{ (\alpha\! -\!1) \left(\alpha ^2\!-2 \alpha
   q\!+q\right)}\!
   +\!\mathcal{O}\!\!\(\!\frac{1}{N_3}\!\)\! \]
\end{equation}
where $q(\alpha)$ is related to $\alpha$ via $\alpha=1-E(q)/K(q)$.
A remarkable feature of this \textit{strong coupling} result is that it resembles a \textit{weak coupling} expansion in $g^2$.
\begin{figure}[t]
\includegraphics[scale=.47]{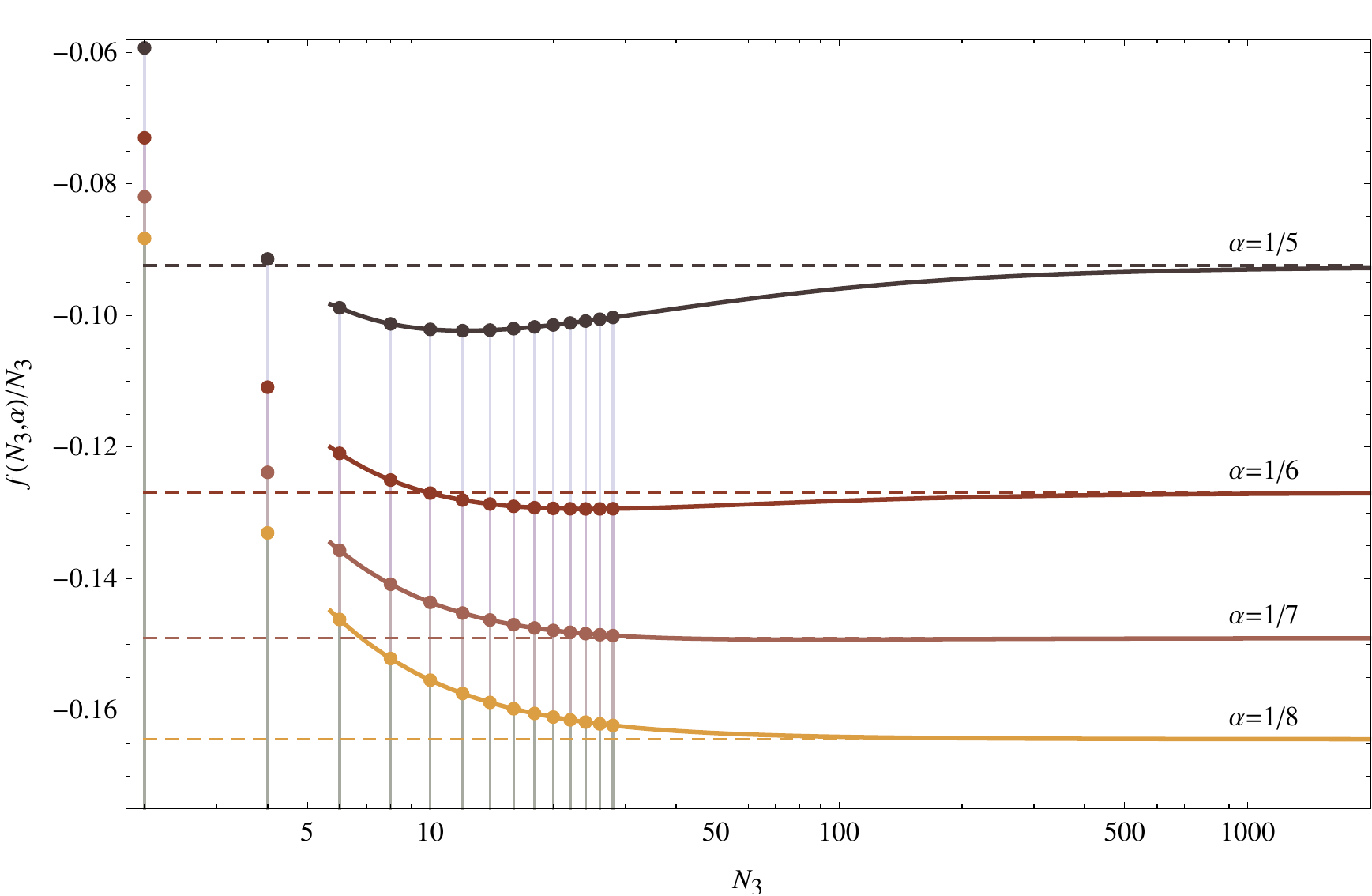}
\caption{
 $f(N_3,\alpha)/N_3$ for several $\alpha$'s as a function of $N_3$. The solid lines are fits in (large) $N_3$. The dashed lines are the strong coupling string theory results
\eq{strongpredic}. The fits asymptote to the dashed lines within the numerical accuracy.
To build this figure we considered in total about $1000$ combinations of three states with up to $56$ magnons and lengths as large as $450$. This computation would be absolutely inconceivable without our main result \eq{eleven}.
} \label{fits}
\end{figure}

To compare with this result we found the corresponding solution
of the two loop Bethe ansatz equations for several
values of $L_1,N_1$ and $N_3$ with very high numerical precision  (see \cite{paper2} for details on the $g^0$ Bethe roots).
Then we plug the Bethe roots into \eq{eleven}
and extrapolate the result to infinite length by increasing $N_1$ and $L_1$ with fixed ratio $\alpha$. We find that the one loop correction (normalized by the tree level result) decays indeed as $f(N_3,\alpha)/L_1^2$ as $L_1$ goes to
infinity.
The values of $f(N_3,\alpha)$ for various $N_3,\alpha$
are shown in fig. \ref{fits}. We observe that $f(N_3,\alpha)$ increases linearly with $N_3$.
To compare with \eq{strongpredic} we found the leading linear term
in $N_3$ with a fit. Note that a priori it is not obvious at all that the weak coupling result (\ref{eleven}) scales as $N_3/L_1^2$. The fact that it does is already highly remarkable and encouraging. Of course, even more striking is the fact that the coefficient matches precisely the string result (\ref{strongpredic})!, see fig. \ref{fits}.

Curiously, at tree level  the weak and strong coupling results match for any finite $N_3$ \cite{paper2}. The numerical analysis at one loop indicates that there is no agreement for finite $N_3$; only the leading term in large $N_3$ matches (\ref{strongpredic}).


\section{Conclusions and Musing}
%
There is a longstanding idea that the complexity of the long-range integrable structure of the AdS/CFT system might come from integrating out some hidden degrees of freedom \cite{Other,Didina}.
The impurities $\theta_j$ and the $\Theta$-derivative realize this idea at weak coupling. Particularly inspiring is the fact that the $\Theta$-derivative not only corrects the states but it also automatically incorporates all one loop Feynman diagrams involved in gluing together the three operators.


As we saw, the $\Theta$-derivative naturally leads  to the Zhukowsky variables. For example the norm $\ckappaa{\theta^{1} }{{\left\<1 |1\right\>}}$ takes the form (\ref{normB}) where in $\phi_k$ we replace \cite{paper4}
\begin{equation}
\prod_{a=1}^{L_1} \frac{u_k-\theta_a^{(1)}+i/2}{u_k-\theta_a^{(1)}-i/2} \to \(\frac{x^+_k}{x^-_k}\)^{L_1} \,.
\end{equation}
This leads to the natural guess that, to all loops, we should simply deform the dispersion and S-matrix in $\phi_k$ as in the spectrum problem. The same comments hold for the main part of the numerator of (\ref{eleven}), the matrix  $G_{nm}$ written in the Appendix.
Hence, with some insight from the spectrum problem, with the help of the $\Theta$-derivative method, and with the inspiration of the Inozemtsev approach
\cite{Didina}, we believe that a conjecture for the all loops structure constants might be within reach for asymptotically large operators.
A first step could be to understand in detail the single magnon case which was so fruitful at one loop. For example, if in (\ref{kappa}) we have
\begin{equation}\la{theta4}
\mathcal{O}(g^4) = \frac{g^4}{8} \sum_{|i-j|\neq 1} \(\partial_{\theta_{i}}-\partial_{\theta_{i+1}}\)^2\!\(\partial_{\theta_{j}}-\partial_{\theta_{j+1}}\)^2 f  +\mathcal{O}(g^6)\nonumber
\end{equation}
then the action of the $\Theta$-derivative on the single magnon state (\ref{coordinateImp}) yields (\ref{coordinate}) up to three-loop order modulo simple boundary terms.
We believe that the same holds for multi-particle states.
Then, a natural conjecture is that \eq{eleven} holds up to two loops. This being investigated \cite{Sau}.
At higher loops, one could try to incorporate the dressing phase
using the boost operator of \cite{BDS}.

\begin{acknowledgments}
We thank D. Serban and A. Sever for very enlightening discussions and suggestions.
We are specially grateful to D. Serban for pointing out the
algebraic description of the Inoszemtsev chain which
was very inspirational.
N.G. (P.V.) would like to thank Nordita and the Perimeter Institute (King's College London, IHP and Nordita) for warm hospitality.  
Research at the Perimeter Institute is supported in part by the Government of Canada through NSERC and by the Province of Ontario through MRI.
\end{acknowledgments}

\appendix

\section{Formulae for Scalar Products}
$$
\vspace{-2.0cm}
$$
\begin{equation}
\<1|1\>=\prod_{m\neq k} \frac{u_k-u_m+i}{u_k-u_m}   \det_{j,k\le N_1} \frac{\partial \phi_k}{\partial u_j}   \la{normB}
\end{equation}
with $e^{i \phi_k}=\prod_{a=1}^{L_1} \frac{u_k-\theta_a^{(1)}+i/2}{u_k-\theta_a^{(1)}-i/2}
\prod_{m\neq k}^{N_1} \frac{u_k-u_m-i}{u_k-u_m+i} $ and similar for $\<2|2\>$.
Finally  \cite{Omar}
$\<1 | \hat{\mathcal{O}}_3| 2 \>={\cal F}
\det\Big([G_{nm}] \oplus [F_{nm}] \Big)$ where  $F_{nm}=\frac{1}{(u_{n}-\theta_m)^2+\frac{1}{4}}$ , \\
$G_{nm}=
\prod_{a=1}^L \frac{v_m-\theta_a^{(1)}+i/2}{v_m-\theta_a^{(1)}-i/2}
\frac{\prod_{k\neq{   n}}^{N_1}(u_k-v_{   m}\!+\! i)}{u_n-v_m}-
\frac{\prod_{k\neq{   n}}^{N_1}(u_k-v_{   m}\!-\! i)}{u_{   n}-v_{   m}} $, \\
${\cal F}=
\frac{
\prod_{m}^{N_3}\prod_{n}^{N_1}(u_n- \theta_m^{(1)}+i/2)/\prod_{m}^{N_3}\prod_{n}^{N_2}(v_n- \theta_m^{(1)}+i/2)
}{
\prod_{n<m}^{N_1}(u_m-u_n)
\prod_{n<m}^{N_2}(v_n-v_m)
\prod_{n<m}^{N_3}( \theta_n^{(1)}-\theta_m^{(1)})
} \,.$


\begin{thebibliography}{99}


\bibitem{paper1}
  J.~Escobedo, N.~Gromov, A.~Sever and P.~Vieira,
  arXiv:1012.2475.


\bibitem{review}
  N.~Beisert {\it et al.},
  Lett.\ Math.\ Phys.\  {\bf 99} (2012) 3,
  arXiv:1012.3982.


\bibitem{BKS}
  J.~A.~Minahan and K.~Zarembo,
  JHEP {\bf 0303} (2003) 013,
  arXiv:hep-th/0212208
  $\bullet$
  N.~Beisert, C.~Kristjansen and M.~Staudacher,
  Nucl.\ Phys.\  B {\bf 664} (2003) 131, arXiv:hep-th/0303060.



\bibitem{paper4}
N.~Gromov and P.~Vieira,
  to appear


\bibitem{roiban}
  R.~Roiban and A.~Volovich,
  JHEP {\bf 0409} (2004) 032
  [arXiv:hep-th/0407140].

\bibitem{Omar}
  O.~Foda,
  arXiv:1111.4663.


%

\bibitem{foot}
In the numerator of (\ref{eleven}) the Bethe roots $v_k$ should be treated as independent of $\theta_j^{(1)}$.

\bibitem{p1}
  K.~Okuyama and L.~S.~Tseng,
  JHEP {\bf 0408} (2004) 055, arXiv:hep-th/0404190 $\bullet$   L.~F.~Alday, J.~R.~David, E.~Gava and K.~S.~Narain,
  JHEP {\bf 0509}, 070 (2005), arXiv:hep-th/0502186.


\bibitem{FT}
  S.~Frolov and A.~A.~Tseytlin,
  Phys.\ Lett.\ B {\bf 570} (2003) 96, hep-th/0306143.


\bibitem{Costa}
  K.~Zarembo,
  JHEP {\bf 1009} (2010) 030, arXiv:1008.1059
$\bullet$   M.~S.~Costa, R.~Monteiro, J.~E.~Santos and D.~Zoakos,
  JHEP {\bf 1011} (2010) 141, arXiv:1008.1070.


\bibitem{paper2}
  J.~Escobedo, N.~Gromov, A.~Sever and P.~Vieira,
  JHEP {\bf 1109} (2011) 029,
  arXiv:1104.5501.


\bibitem{strange}
See   A.~Bissi, T.~Harmark and M.~Orselli,
  arXiv:1112.5075 for an interesting effort at obtaining a one loop/strong coupling match using the coherent state formalism proposed in \cite{paper2}.

\bibitem{Other}
  A.~Rej, D.~Serban and M.~Staudacher,
  JHEP {\bf 0603} (2006) 018,
  arXiv:hep-th/0512077
  $\bullet$  N.~Gromov and V.~Kazakov,
  Nucl.\ Phys.\  B {\bf 780} (2007) 143,
  arXiv:hep-th/0605026 $\bullet$  N.~Gromov and P.~Vieira,
  Nucl.\ Phys.\  B {\bf 790} (2008) 72,
  arXiv:hep-th/0703266.


\bibitem{Didina}
  D.~Serban and M.~Staudacher,
  JHEP {\bf 0406} (2004) 001, arXiv:hep-th/0401057

\bibitem{BDS}
  N.~Beisert, V.~Dippel and M.~Staudacher,
  JHEP {\bf 0407} (2004) 075
  [arXiv:hep-th/0405001]. $\bullet$  T.~Bargheer, N.~Beisert and F.~Loebbert,
  J.\ Phys.\ A  {\bf 42} (2009) 285205
  [arXiv:0902.0956 [hep-th]].

\bibitem{Sau}
N.~Gromov. S.~Valatka, P.~Vieira, in progress

\end{thebibliography}
\end{document}